\documentclass[%
 aip, jap, amsmath,amssymb,
 preprintnumbers,%
 reprint,%
]{revtex4-2} 
\usepackage{booktabs}
\DeclareMathAlphabet{\mathcalligra}{T1}{calligra}{m}{n}
\usepackage[english]{babel}
\usepackage{amsmath}
\usepackage{bbm}
\usepackage{graphicx}

\usepackage{comment}
\usepackage{multirow}
\usepackage[dvipsnames]{xcolor}
\usepackage{bm}
\usepackage{mathtools}
\usepackage{amsmath}
\usepackage{rotating}
\definecolor{oceanboatblue}{rgb}{0.0, 0.47, 0.75}
\definecolor{orange}{rgb}{1,0.5,0}
\definecolor{goodgreen}{rgb}{0.1,0.5,0}
\definecolor{goodred}{rgb}{0.7,0,0}
\usepackage[colorlinks,urlcolor=goodgreen,citecolor=blue,linkcolor=goodred]{hyperref}
\usepackage{cleveref}
\usepackage{tikz} 
\usepackage{bbm,amsfonts}
\usepackage{tocvsec2}
\usepackage{enumitem} 
\usepackage{float} 

\usepackage{cancel} 

\makeatletter 
\newcommand{\ii}{\text{i}}
\newcommand{\ee}{\text{e}}
\renewcommand{\exp}[1]{\ee^{#1}}

\usepackage[normalem]{ulem}
\usepackage{soul}

\usepackage{braket}

\newcommand{\epsA}{\epsilon_{\rm A}}
\newcommand{\epsB}{\epsilon_{\rm B}}

\newcommand{\bk}{\boldsymbol{k}}
\newcommand{\Hp}{\mathcal{H}(\bk)}
\newcommand{\Hm}{\mathcal{H}(-\bk)}

\newcommand{\HpC}{\mathcal{H}^*(\bk)}

\newcommand{\HpD}{\mathcal{H}^\dagger(\bk)}
\newcommand{\HmD}{\mathcal{H}^\dagger(-\bk)}

\newcommand{\HpT}{\mathcal{H}^\mathrm{T}(\bk)}

\newcommand{\Inv}{\mathcal{I}}
\newcommand{\PT}{\mathcal{PT}_+}
\newcommand{\Parity}{\mathcal{P}}

\usepackage{lipsum}

\begin{document}

\author{C. Mart\'{\i}nez-Strasser}
\email{cmartinez089@ikasle.ehu.eus}

\affiliation{Donostia International Physics Center (DIPC), Manuel de Lardizbal 4, 20018 San Sebasti\'an, Spain}
\affiliation{Advanced Polymers and Materials: Physics, Chemistry and Technology, Chemistry Faculty (UPV/EHU), Paseo M. Lardizabal 3, 20018 San Sebastian, Spain}

\author{D. Bercioux}
\email{dario.bercioux@dipc.org}
\affiliation{Donostia International Physics Center (DIPC), Manuel de Lardizbal 4, 20018 San Sebasti\'an, Spain}
\affiliation{IKERBASQUE, Basque Foundation for Science, Plaza Euskadi 5 48009 Bilbao, Spain}

\author{N. Leumer}
\email{nico.leumer@pwr.edu.pl}
\affiliation{Department of Theoretical  Physics, Wrocław University of Science and Technology, Wybrzeże Wyspiańskiego 27, 50-370 Wrocław, Poland}

\title{Tuning topological phases and exceptional points in a non-Hermitian Rice Mele model beyond nearest neighbors}

\begin{abstract}
Exceptional points are degeneracies characteristic of non-Hermitian operators, where eigenvalues and eigenvectors coalesce, rendering the Hamiltonian defective. We investigate the exceptional-point structure and topological properties of a generalized non-Hermitian Rice-Mele model with balanced gain and loss, as well as next-nearest-neighbor hopping. The system hosts only second-order exceptional points under both periodic and open boundary conditions. Under periodic boundary conditions, the exceptional points in parameter space lie on lines and ellipses that are independent of the next-nearest-neighbor hopping, since the latter enters the bulk Hamiltonian only as an identity contribution. Under open boundary conditions, this independence is broken: the next-nearest-neighbor hopping not only shifts the energy of existing exceptional points but also generates new ones, with a specific condition signaling a topological gap closing observed only in the open-boundary spectrum. At special parameter points, multiple simultaneous second-order exceptional points yield degenerate configurations whose degeneracy grows with system size. Exceptional point locations are identified numerically via the condition number of the eigenvector matrix and confirmed by Jordan decomposition. The topological phase diagram, computed via a winding number framework for non-Hermitian systems without symmetry protection, reveals sectors with zero, one, and two edge states; the bulk-boundary correspondence is confirmed, and the non-Hermitian skin effect is absent.  
\end{abstract}

\maketitle
\section{Introduction}
Exceptional points (EPs) are singular degeneracies unique to non-Hermitian (nH) operators, at which two or more eigenstates coalesce into a single one, leaving the Hamiltonian defective.~\cite{bergholtz2021, heiss2004, Heiss2012, kato1966perturbation} While they arise naturally in intrinsically non-Hermitian Hamiltonians, they can also emerge in effective non-Hermitian descriptions of Hermitian systems, such as dynamical matrices derived from bosonic quadratic Hamiltonians.~\cite{Wang2019, Flynn2020, Perina2022, Wakefield2024} The emergence of EPs generally requires fine-tuning of system parameters, rendering their existence intimately tied to the underlying symmetries of the system, particularly the existence of EPs of higher order.~\cite{sayyad2022realizing} A hallmark of $n$-th order EPs is their potential for enhanced sensing, since small perturbations $\epsilon$ cause a magnified response\cite{Grom2025, wiersig2014, Liu2016} $\epsilon^{1/n}$ and which has been observed experimentally in optical micro cavities.~\cite{chen2017} The Petermann factor quantifies eigenstate non-orthogonality in non-Hermitian systems and diverges at EPs, where right and left eigenmodes become mutually orthogonal: this divergence signals the breakdown of standard mode orthogonality and leads, \emph{e.g.}, to a broadening of the laser linewidth beyond the Schawlow-Townes limit.~\cite{berry2003} However, the accompanying noise enhancement cancels the sensing gain at the EP, leaving the signal-to-noise ratio unchanged.~\cite{wang2020, Loughlin2024, Langbein2018, Zhang2019} Beyond their spectral properties, EPs are indeed topological in nature and adiabatically encircling an EP in parameter space leads to a non-trivial permutation of eigenvalues and eigenstates — they do not return to themselves after a full loop as established theoretically\cite{heiss2001chirality, bergholtz2021, kawabata2019symmetry,yang2024homotopy} and confirmed experimentally.~\cite{dembowski2001, dembowski2004} Since transposition and Hermitian conjugation are inequivalent, the notion of non-Hermiticity (nHy) extends the well-established Altland-Zirnbauer (AZ) classification,~\cite{altland1997nonstandard,
schnyder2008classification} to an extended scheme of 38-symmetry classes.~\cite{kawabata2019symmetry}

The discovery of the non-Hermitian skin effect (NHSE), among other phenomena, revealed that standard topological invariants and the bulk-boundary correspondence require generalization in nH systems both for dispersive bands and for flat ones.~\cite{yao2018edge,
zhong2025topological, leumer2025,Lee2019,Zhang2020, kunst2018biorthogonal, okuma2020topological, song2019non,Martinez_Strasser_2023,Martinez_Strasser_2024,Esparza_2026} Famously, the conventional ‘real’ Berry phase successfully predicts gapless modes in Hermitian systems, while it generally fails in nH models,~\cite{Lieu2018} and complex generalizations based on a bi-orthogonal definition have been proposed.~\cite{Dattoli1990, Garrison1988, Mostafazadeh1999} In recent years, the community proposed a growing collection of techniques, \emph{e.g.}, a bi-orthogonal approach to the bulk-boundary correspondence defining the bi-orthogonal polarization,~\cite{kunst2018biorthogonal} the concept of a generalized Brillouin-zone~\cite{Yokomizo2019, Zhang2020}(GBZ), and generalized non-Bloch or spectral winding numbers\cite{yao2018edge, okuma2020topological, Gong2018, kawabata2019symmetry, zhong2025topological, Zhang2020} to define the topological phase diagram from the bulk Hamiltonian and to predict the existence of edge states under open boundaries conditions. Particularly, Ref.~[\onlinecite{zhong2025topological}] defines a winding number that sets the amount of edge states in one-dimensional (1D) systems that support non-zero energy edge states but lack spatial symmetries.

In this work, we consider a non-Hermitian generalization of the  Rice-Mele (RM) model;~\cite{Rice1982} this is an extension of the Su-Schrieffer-Heeger (SSH) model.~\cite{Su1979} It features independent sublattice onsite energies and next-nearest-neighbor (nnn) hopping, extended to the non-Hermitian regime through gain and loss ($\pm i\gamma$) on the two sublattices. With the exception of time-reversal symmetry, the interplay of all parameters causes the absence of all symmetries, particularly chiral and particle-hole symmetries. Based on the winding number from Ref.~[\onlinecite{zhong2025topological}], the bulk Hamiltonian predicts the existence of finite-energy edge states under open-boundary conditions (OBC). We discuss the phase diagram and how the localization has been modified by the present nnn hopping. Still, our main focus rests on the EPs. Generally, we find EPs of the second order only, both under PBC and OBC. However, parameter constraints for EPs highly differ between periodic-boundary conditions (PBC) and OBC, with the nnn hopping having no effect on the bulk EPs but a strong effect on the OBC EPs. We further find that at specific parameter points multiple second-order EPs (EP2s) occur simultaneously --- one for each coalescing eigenvalue pair --- yielding a degenerate EP$_2$ configuration whose degeneracy grows with the system size $N$.

Variants of the present model have appeared in the literature in different contexts, primarily in relation to $\mathcal{PT}$-symmetry, and are thus distinct from the results reported here. In Ref.~[\onlinecite{longhi2013convective}], an identical model is considered, with the focus placed on $\mathcal{PT}$-symmetry breaking via wave-packet dynamics, while Ref.~[\onlinecite{li2020}] investigates the breaking of $\mathcal{PT}$-symmetry in an SSH chain where onsite gain and loss are confined to the terminal sites. This line of inquiry is extended in Ref.~[\onlinecite{xing2017}], where gain and loss terms are permitted at arbitrary positions along the chain without the constraint of sublattice filling. For generic boundary conditions, Ref.~[\onlinecite{ghosh2025edge}] investigates edge states and persistent currents in an SSH chain extended by next-nearest-neighbor hopping, subjected to a magnetic flux and incorporating onsite gain and loss terms. 

Reference~[\onlinecite{lai2025non}] employs a variant of the present model incorporating nnn hopping restricted to a single sublattice, where both the adaptation to a nonreciprocal coupling and the inclusion of balanced gain and loss are required for the NHSE to manifest. Similarly, Ref.~[\onlinecite{jiang2024}] employs a limiting case of our model, with absent onsite energies and imaginary nnn hopping, to demonstrate the NHSE's tunability by gain and loss terms. Lastly, Ref.~[\onlinecite{wu2022}] extends an SSH chain by longer, next-next nearest neighbor hopping, and studies the topological transitions induced by onsite/gain loss.

The manuscript is organized as follows. In Sec.~\ref{sec:model} we introduce the model and discuss its symmetries. Section~\ref{sec:EPs} focuses on exceptional points with dedicated parts for the bulk constraints, under periodic or open boundary conditions, respectively. Based on the discussion on symmetries, we use a winding number invariant for nH systems to topologically classify our model and to obtain the topological phase diagram to uncover the presence of edge states in Sec.~\ref{sec: edge states}. Finally, we conclude our results in section~\ref{sec: conclusion}. A few technical Appendices are included at the end of the manuscript.

\section{Model}
\label{sec:model}
%
%
\begin{figure}[!t]
    \centering
    \includegraphics[width=0.85\columnwidth]{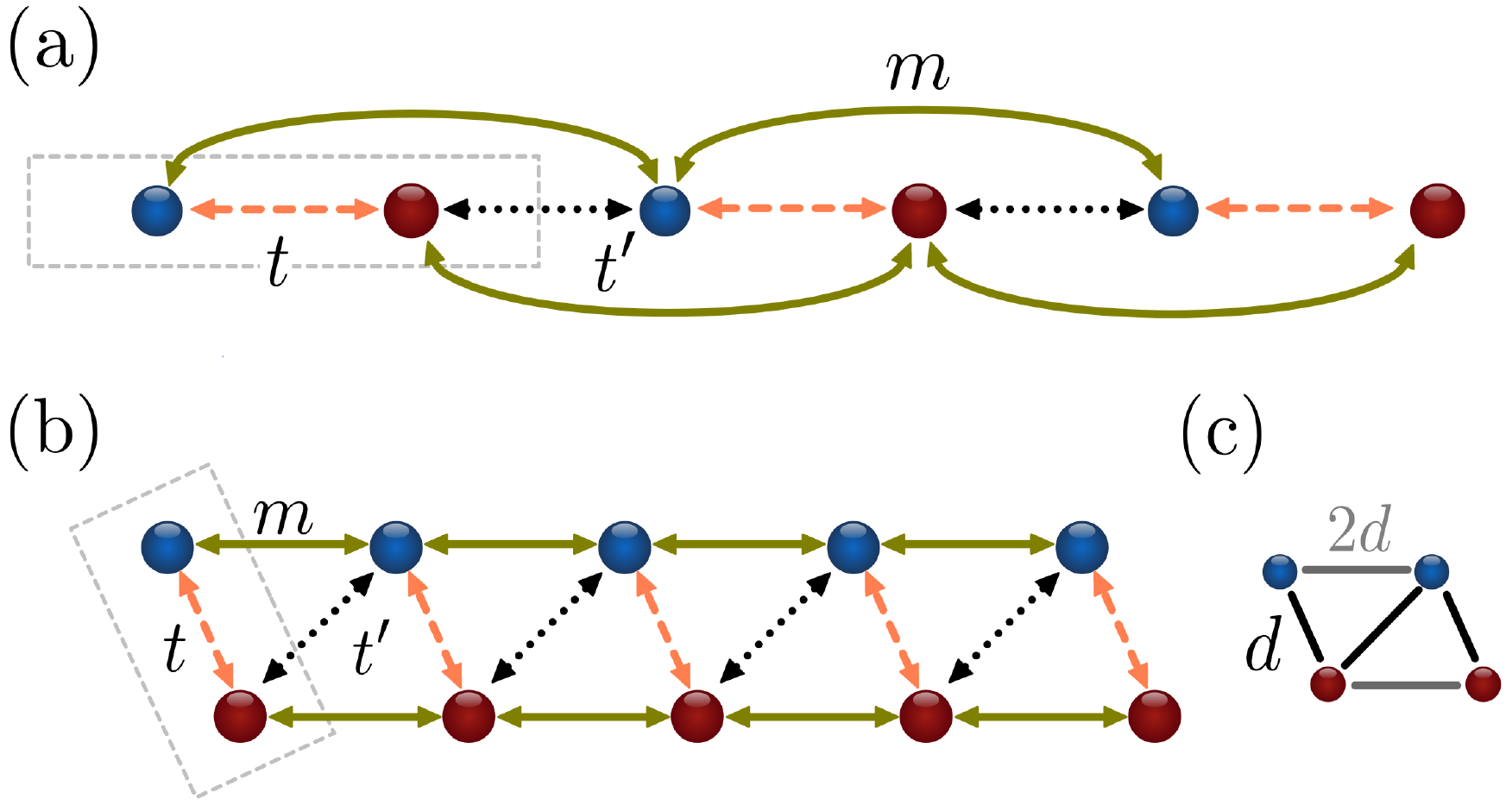}
    \vspace{-0.5cm}
    \caption{\textbf{Sketch of the model in real space.} (a) SSH chain with $t$, $t'$ for intra- and inter-cell hopping, and $m$ for long-range hopping terms. Sublattice A (B) sites, shown in blue (red), encode gain/loss $\pm \ii \gamma $ and on-site energies $\epsilon_{\rm A, B}$. The gray box indicates the unit cell. (b) The system is represented as a quantum ladder, with sublattices coupled by $t$ and $t'$. (c) Real space distances of panel (b) in terms of the lattice constant $d$.} 
    \label{Fig1}
\end{figure}
%
%
%
We consider the 1D chain illustrated in Fig.~\ref{Fig1}(a) and described by the following tight-binding model
%
%
    \begin{align}\label{Ham_TB}
    H&=\sum_{n=1}^N(\epsA+\ii \gamma)c_{\text{A},n}^\dag c_{\text{A},n}+(\epsB-\ii \gamma) c_{\text{B},n}^\dag c_{\text{B},n}) \nonumber+ \\
    &\quad -\left[t~ c_{\text{A},n}^\dag c_{\text{B},n} +t' c_{\text{B},n}^\dag c_{\text{A},n+1} + \right. \\
    &\quad m \left.\left(c_{\text{A},n}^\dag c_{\text{A},n+1}+c_{\text{B},n}^\dag c_{\text{B},n+1}\right) + \text{h.c.}\right]. \nonumber
\end{align}
%
%
It consists of two distinct sublattices A, B, and where $t$, $t'$ denote intra- and intercell hoppings of an SSH chain,~\cite{Su1979} and $\epsA$, $\epsB$ abbreviate onsite energies. Together, they constitute a RM~\cite{Rice1982} model that is extended by the nnn hopping $m$~\cite{Perez2018,Perez2019} and turns nH due to onsite gain/loss terms $\pm \ii \gamma$.~\cite{Zhu2014} In Eq.~\eqref{Ham_TB}, $c_{s,n}^\dag$ 
creates a spinless excitations on site $s,n$ with $s = \text{A, B}$. Note that all quantities introduced in Eq.~\eqref{Ham_TB} are real and that we adopt $2\epsilon_\pm  = \epsA \pm \epsB$ hereinafter.

%
%
\begin{table*}[!ht]
\centering
\setlength{\tabcolsep}{22pt}
\begin{tabular}{l l l l}
\hline
Symmetry &  Symmetry constraint & Matrix & Parameter constraint \\
\hline

TRS II ($\rm TRS^\dagger$)
&
$\Hm =  \mathcal{C}_+ \HpT \mathcal{C}_+^\dagger$
& $\mathcal{C}_+ =\mathbbm{1}_2$
&Always
\\
TRS I 
&
$\Hm =  \mathcal{T}_+ \HpC \mathcal{T}_+^\dagger$
&$\mathcal{T}_+ = \mathbbm{1}_2$
&$\gamma = 0$
\\
$\Inv$
&
$\HmD =  \Inv \Hp \Inv^{-1}$
& $\Inv=\sigma_x$
&$\epsilon_-=0$
\\
$\Parity$
&
$\Hm =  \Parity \Hp \Parity^{-1}$
&$\Parity=\sigma_x$
&$\epsilon_-=\gamma=0$
\\
$\mathcal{PT}$
&
$\Hp =  (\PT) \HpC (\PT)^{-1}$
& $\PT=\sigma_x \hat{K}$
&$t'=\gamma=\epsilon_-=0$
\\
psH
&
$\Hp =  \varsigma \HpD \varsigma^{-1}$
& $\varsigma=\mathbbm{1}_2$
& $\gamma=0$
\\
CS
&
$\Hp =  -\Gamma \HpD \Gamma^{-1}$
&$\Gamma = \sigma_z$
&$\epsilon_\pm=m=0$
\\
SLS
&
$\Hp =  -\mathcal{S} \Hp \mathcal{S}^{-1}$
&$\mathcal{S}  = \sigma_z$
&$\epsilon_\pm=\gamma=m=0$
\\
psCS
&
$\HpT =  -\Lambda \Hp \Lambda^{-1}$
&$\Lambda = -\ii \sigma_y$
&$\epsilon_+ = m=0$
\\
PHS I 
&
$\Hm = - \mathcal{C}_- \HpT \mathcal{C}_-^\dagger$
&$\mathcal{C}_- = \sigma_z$
&$\epsilon_\pm =\gamma=m=0$
\\
PHS II ($\rm PHS^\dagger$)
&
$\Hm = - \mathcal{T}_- \HpC \mathcal{T}_-^\dagger$
&$\mathcal{T}_- = \sigma_z$
&$\epsilon_\pm =m=0$
\\
$\mathcal{CP}$
&
$\Hp =  -(\mathcal{CP}) \HpC (\mathcal{CP})^{-1}$
&$\mathcal{CP} = \ii \sigma_y\hat{K}$
&$\epsilon_+=t'=m=0$
\\
\hline
\end{tabular}
\caption{Symmetries of the non-Hermitian bulk Hamiltonian $\mathcal{H}(k)$~\eqref{eq: Full Ham in kspace}
    and their parameter constraints, classified according to the 38-fold
    non-Hermitian symmetry scheme of Kawabata et al.~\cite{kawabata2019symmetry}
    using the notation of Sayyad--Kunst.~\cite{sayyad2022realizing} Unitary matrices $A\in \{\Gamma, \Lambda,\varsigma, \mathcal{S}, \Parity, \Inv\}$ ($A\in \{\mathcal{C}_\pm, \mathcal{T}_\pm\}$ ) satisfy $A^2 = \mathbbm{1}$ ($AA^* = \xi_A \mathbbm{1}$, $\xi_A = \pm 1$).
    \emph{Notation}: 1) $\sigma_{x,y,z}$: Pauli matrices, 2) $\hat{K}$: operator of complex conjugation, 3) $\mathbbm{1}_2$ Identity matrix of size $2$. \emph{Acronyms}: 1) TRS: time reversal sym., 2) $\mathcal{I}$: inversion sym., 3) $\Parity$: parity, 4) $\PT$: parity time-reversal sym., 5) psH: pseudo Hermitian sym., 6) CS: chiral sym., 7) SLS: sublattice sym., 8) psCS: pseudochiral sym., 9) PHS: particle hole sym., 9) $\mathcal{CP}$: parity-particle-holy sym.
     }
  \label{tab:symmetries}\end{table*}
%
%
From a Fourier analysis of the Hamiltonian $H = \sum_k \psi_k^\dagger \mathcal{H}(k) \psi_k$, $\psi_k = (c_{\text{A},k},c_{\text{B},k})^\mathrm{T}$, we find 
\begin{align}\label{eq: Full Ham in kspace}
\mathcal{H}(k)=[ \epsilon_+ -2m\cos(k) ]\mathbbm{1}_2 + \mathcal{H}_{\rm RM}(k)
\end{align}
%
%
where
%
%
\begin{align}\label{eq: RM Ham in kspace}
    \mathcal{H}_{\rm RM}(k) &= \begin{pmatrix}
        \Gamma & - t-t'e^{-\ii k}\\
         - t-t'e^{\ii k} & -\Gamma
    \end{pmatrix}
\end{align}
%
%

\noindent is a nH RM Hamiltonian with complex $\Gamma = \epsilon_- + \ii \gamma$ and the lattice constant $d$ being absorbed into $k$. As expected, the terms $t$, $t'$ yield the known off-diagonal elements for an SSH chain since $t$, $t'$ operate between the sublattices as illustrated in Fig.~\ref{Fig1}(b). Onsite gain/loss terms $\pm \ii \gamma$ are shifted by $\pm \epsilon_-$ respectively, while the average energy $\epsilon_+$ is a global shift. Since the nnn hopping term $m$ couples sites within each sublattice equally, its contribution $-2m\cos(k)$ is proportional to the identity matrix $\mathbbm{1}_2$.~\cite{Perez2018,Perez2019}

Diagonalizing the bulk Hamiltonian $\mathcal{H}(k)$ gives the dispersion relation $E_\pm (k) = \epsilon_+ -2m\cos(k) +E_{\rm RM, \pm} (k) $ with complex RM energies
\begin{align}\label{eq: RM dispersion relation}
    E_{\rm RM, \pm} (k) = \pm \sqrt{\Gamma^2 + t^2+(t')^2 + 2tt'\cos(k)}.
\end{align}
Generally, $E_{\rm RM, +}$, $E_{\rm RM, -}$ are \emph{not} complex conjugates yet $E_{\rm RM, \pm}$ is symmetric in $k$ which originates from the present/absent bulk symmetries.

\subsection{Bulk symmetries}

The Altland-Zirnbauer (AZ) scheme~\cite{altland1997nonstandard,
schnyder2008classification} is no longer valid for nH systems, given that transposition and complex conjugation are generally distinct, which naturally doubles the number of symmetry operators.  Additionally, for nH systems, the sublattice symmetry operator~(SLS) is decoupled from the chiral symmetry~(CS) operator. Consequently, after combining with all four anti-unitary generators, it gives rise to 22 independent classes, which overall result in 38-fold symmetry classes of
Kawabata \textit{et al.}~\cite{kawabata2019symmetry} after considering unifying symmetry operations. 

Adopting the notation of Ref.~[\onlinecite{sayyad2022realizing}], we list in Table~\ref{tab:symmetries}  each symmetry, its definition, and the parameter constraints required for its realization. Generally, only the time-reversal symmetry TRS\,II $C_+\hat{K}$ is preserved with $C_+ = \mathbbm{1}_2$, where $\hat{K}$ is the complex conjugation operator, since time reversal symmetry preserves the sublattices $A$, $B$ that form the basis of Eq.~\eqref{eq: RM Ham in kspace} in agreement with Refs.~[\onlinecite{kawabata2019symmetry, bergholtz2021exceptional, sayyad2022realizing}]. 

Physically, the preservation of TRS\,II reflects the absence of non-reciprocity,~\cite{okuma2020topological} \emph{i.e.}, left/right moving solutions correspond to the same energy $E_\pm (k)=E_\pm (-k)$ as manifests in $\mathcal{H}_{12}(-k) = \mathcal{H}_{21}(k)$ and the absence of the NHSE. In contrast, the nH generalization of the standard anti-unitary time-reversal of the AZ scheme~\cite{altland1997nonstandard, schnyder2008classification} TRS (TRS\,I in
Ref.~[\onlinecite{sayyad2022realizing}]) is $\mathcal{T}_+ \hat{K}$ with $\mathcal{T}_+ = \mathbbm{1}_2$ but requires that $\gamma=0$. Note that in the Hermitian limit only, TRS and TRS$^\dagger$ coincide with the AZ time-reversal symmetry.~\cite{altland1997nonstandard}

Similarly, the non-Hermiticity separates parity $\mathcal{P}$ and spatial inversion $\mathcal{I}$. While both operations exchange the sublattices, in our model, $\mathcal{P} = \mathcal{I} = \sigma_x$, spatial inversion $\mathcal{I}$ is preserved whenever $\epsilon_- = 0$, i.e., $\epsilon_A = \epsilon_B$, whereas parity demands $\epsilon_- = \gamma = 0$. Parity-time symmetry $\mathcal{PT}=\sigma_x \hat{K}$ combines $\mathcal{P}$ and TRS I. Since $\mathcal{PT}$ relates $\mathcal{H}(k)$, $\mathcal{H}^*(k)$, $\mathcal{PT}$ demands $\epsilon_- = \gamma = 0$ but also that $t'=0$. The last symmetry without constraints on $m$ is pseudo-Hermiticity (psH) with $\varsigma = \mathbbm{1}_2$ demanding that $\gamma=0$, i.e., an Hermitian system. 

In the absence of nnn hopping $m=0$, the Hamiltonian reduces to an nH Rice-Mele chain, which permits additional symmetries, most notably chiral and particle-hole (PHS) symmetries. Although both CS and SLS are represented by the same operator $\sigma_z$ in the Hermitian limit, non-Hermiticity separates them into distinct symmetries. Specifically, CS ($\Gamma = \sigma_z$) requires $\epsilon_\pm = m = 0$
, while SLS ($\mathcal{S} = \sigma_z$) imposes the more restrictive condition $\gamma = \epsilon_\pm = m = 0$. Without AZ counterpart, nHy allows also for pseudochiral symmetry (psCS) $\Lambda = -\ii\sigma_y$ whenever $\epsilon_- = m=0$. Regarding particle-hole symmetry PHS I (PHS II), we find $\mathcal{C}_- = \sigma_z$ ($\mathcal{T}_- = \sigma_z$) and its restoration requires $m=\epsilon_\pm =\gamma =0$ ($m=\epsilon_\pm=0$). Lastly, we find parity-particle-holy symmetry $\mathcal{CP}=\ii \sigma_y \,\hat{K}$ under the constraints that $t'=m=\epsilon_+=0$.

Based on the bulk symmetries, we calculate the winding number in section~\ref{sec: edge states} proposed by Ref.~[\onlinecite{zhong2025topological}] for nH systems, which manifests the presence of edge states in the model.

\section{Exceptional points}
\label{sec:EPs}
\subsection{Bulk exceptional points}\label{EP PBC}
A hallmark of nH systems is the occurrence of EPs~| points in parameter space where the Hamiltonian becomes defective, and two or more eigenvalues coalesce together with their associated eigenvectors. Here, we investigate the EPs of the bulk Hamiltonian~\eqref{eq: Full Ham in kspace}.

To obtain the EPs, we will verify when the discriminant of the secular equation associated with this Hamiltonian is zero. Specifically, if we express this Hamiltonian as $\mathcal{H}(k) = d_0(k)\,\mathbbm{1}_2 + \bm{d}(k)\cdot\bm{\sigma}$, where $\bm{d}(k)=\bm{d}_\text{R}(k)+\mathrm{i}\,\bm{d}_\text{I}(k)$ and $\bm{\sigma}=(\sigma_1,\sigma_2,\sigma_3)$ is the vector of Pauli matrices, the mathematical condition for obtaining EPs is given by the solution of the following set of equations: $|\bm{d}_\text{R}|^2-|\bm{d}_\text{I}|^2=0$ and $\bm{d}_\text{R}\cdot\bm{d}_\text{I}=0$.~\cite{bergholtz2021} Concretely, this set of equations reads
%
%
\begin{subequations}\label{eq EPs}
\begin{align}
    & t^2+(t')^2+2t t' \cos(k d)+(\epsilon_-^2-\gamma^2)=0\,, \label{eq EP 1}\\
    & \epsilon_- \gamma=0, \label{eq EP 2}
\end{align}
\end{subequations}
%
%
\emph{i.e.}, the gap term between the bands from Eq.~\eqref{eq: RM dispersion relation} vanishes and the eigenvectors coalesce, cf. Appendix~\ref{app: Coalescence of bulk eigenstates}. Notably,  Eq.~\eqref{eq EP 2} implies $\epsilon_-=0$, $\gamma\neq0$ as the sole possibility since $\gamma=0$ restores Hermiticity; thus, indicating the absence of EPs for our choice of real parameters. 

Since the bulk Hamiltonian~\eqref{eq: Full Ham in kspace} is $2\times 2$, our model necessarily hosts EP's of order two only as $\mathcal{H}(k_\text{EP})$ is defective with a single $2\times 2$ Jordan block. This result remains actually true even for OBC and finite size, as we show in Sec.~\ref{EP for OBC}.

Unlike the EP's order, the boundary condition changes the parameter constraints. With both $m$, $\epsilon_+$ absent in Eqs.~\eqref{eq EPs}, they do not influence bulk EPs | however | we show the striking difference in Sec.~\ref{EP for OBC} where the nnn hopping term $m$ impacts the parameter condition and allows the tuning of degenerate EPs.

Nonetheless, the non-real eigenvalues associated to bulk EPs depend on $\epsilon_+$, $m$, the latter moderating between purely imaginary or genuinely complex ones. Solving Eq.~\eqref{eq EP 1} w.r.t. $\gamma$, we find critical values
%
%
\begin{figure}
    \centering
    \includegraphics[width=\columnwidth]{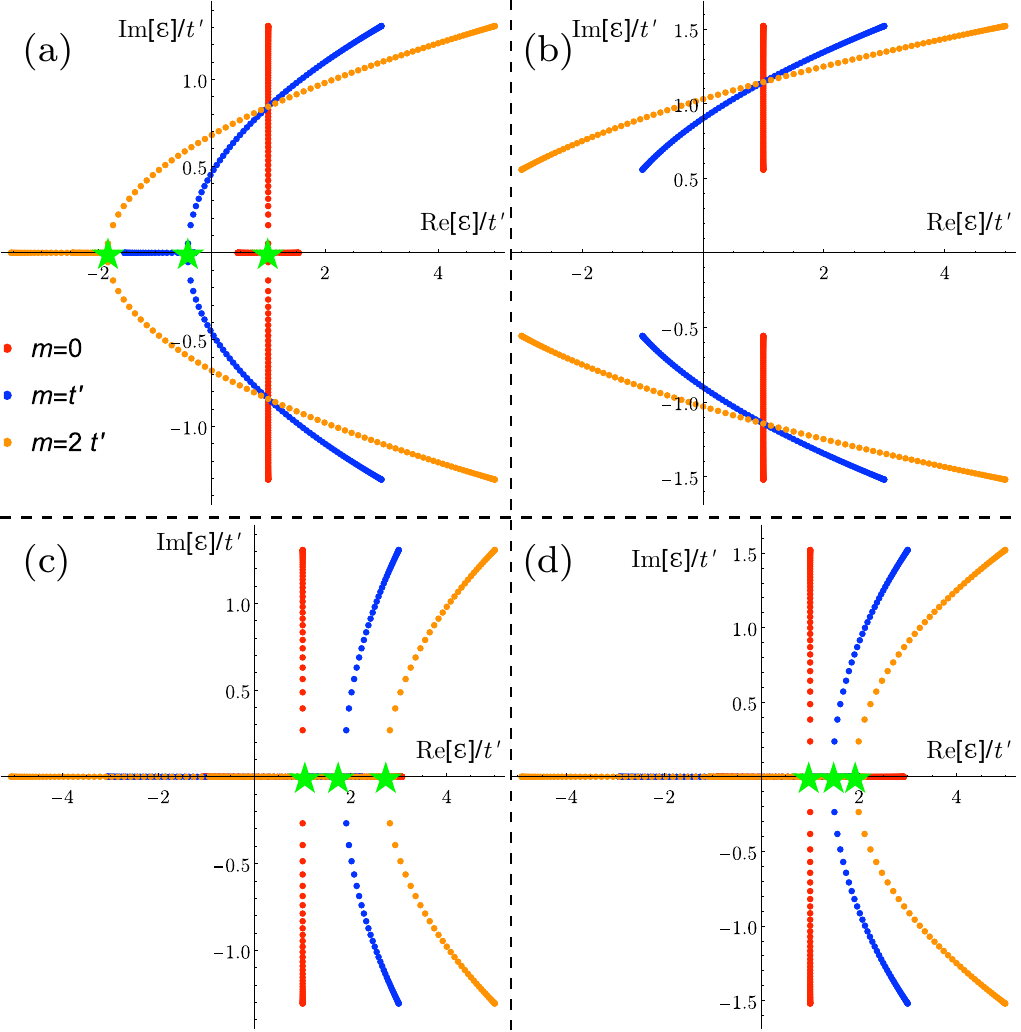}
\caption{\textbf{Bulk complex spectrum and exceptional points.} Bulk energy spectrum of $\mathcal{H}(k)$ in the complex plane, parametrized by $k \in \text{BZ}$, for fixed $\epsilon_+ = t'$ and $\epsilon_- = 0$. Each panel shows three superposed spectra corresponding to $m = 0$ (red), $m = t'$ (blue), and $m = 2t'$ (orange). Parameters: (a) $t = t'/2$, $\gamma = 1.4\,t'$; (b) $t = t'/2$, $\gamma = 1.6\,t'$; (c) $t = 3t'/2$, $\gamma = 1.4\,t'$; (d) $t = 3t'/2$, $\gamma = 1.6\,t'$. For $m = 0$, the spectrum is purely imaginary (up to the constant real shift $\epsilon_+$), while for $m \neq 0$, the next-nearest-neighbor hopping introduces a $k$-dependent dispersion of the real part, producing genuinely complex bands. Green stars mark the bulk EPs, which lie on the real axis at $\mathrm{Re}\,(E_\text{EP}) = \epsilon_+ - 2m\cos(k_\text{EP} d)$. Panel~(b) lies in regime~III ($\gamma > t + t'$) and consequently exhibits no EPs.}
    \label{EP_bulk}
\end{figure}
%
%
%
%
\begin{align}\label{gamma_critical}
    \gamma_\text{c}(k)=\pm\sqrt{t^2+t'^2+2tt'\cos(k)}
\end{align}
%
%
that produce a change in the nature of the spectrum and~| assuming $t,t'>0$ for definiteness | its $k$-dependence defines three distinct regimes:
%
%
\begin{itemize}
    \item \emph{Regime I}: For $\gamma<|t-t'|$, the condition has no solution in the BZ and the spectrum is real for all $k$.
    \item \emph{Regime II}: For $|t-t'|<\gamma<t+t'$, the condition is met at a pair of momenta $\pm k_\text{EP}$ with $$\cos(k_\text{EP} d)=\frac{\gamma^2-t^2-t'^2}{2tt'}.$$ These are the bulk EPs; the spectrum is \emph{non-real} inside the interval $[-k_\text{EP},k_\text{EP}]$ and real outside.
    \item \emph{Regime III}: For $\gamma>t+t'$, the EP condition is exceeded throughout the BZ and the spectrum is non-real for all $k$.
\end{itemize}
%
%
In the non-real regime, $\gamma>\gamma_\text{c}(k)$, the eigenvalues read
%
%
\begin{align}\label{EP equation transition}
    E_\pm(k) = \epsilon_+ - 2m\cos(k) \pm \mathrm{i}\sqrt{\gamma^2 - \gamma_c^2(k)},
\end{align}
%
%
which makes the role of $\epsilon_+$ and $m$ transparent: for $m=0$ the real part is flat and equal to $\epsilon_+$ (purely imaginary eigenvalues if $\epsilon_+=0$ as well), while for $m\neq 0$ the nnn hopping gives the real part a $k$-dependent dispersion, yielding a genuinely complex spectrum. A noteworthy special case is $t=t'$, where the lower threshold $|t-t'|$ collapses to zero: any infinitesimal $\gamma$ produces an EP at $k=\pi/d$, and there is no $\gamma>0$ regime with a fully real spectrum. We summarize all these results in the various panels of Fig.~\ref{EP_bulk}. To conclude, we have shown that the existence of the EPs and their momentum $k_\text{EP}$ are independent of $\epsilon_+$ and $m$; their position in the complex energy plane, however, does depend on both parameters.
\subsection{Parameter dependence of EPs under PBC and finite size}
%
%
\begin{figure*}
    \centering
    \includegraphics[width=\textwidth]{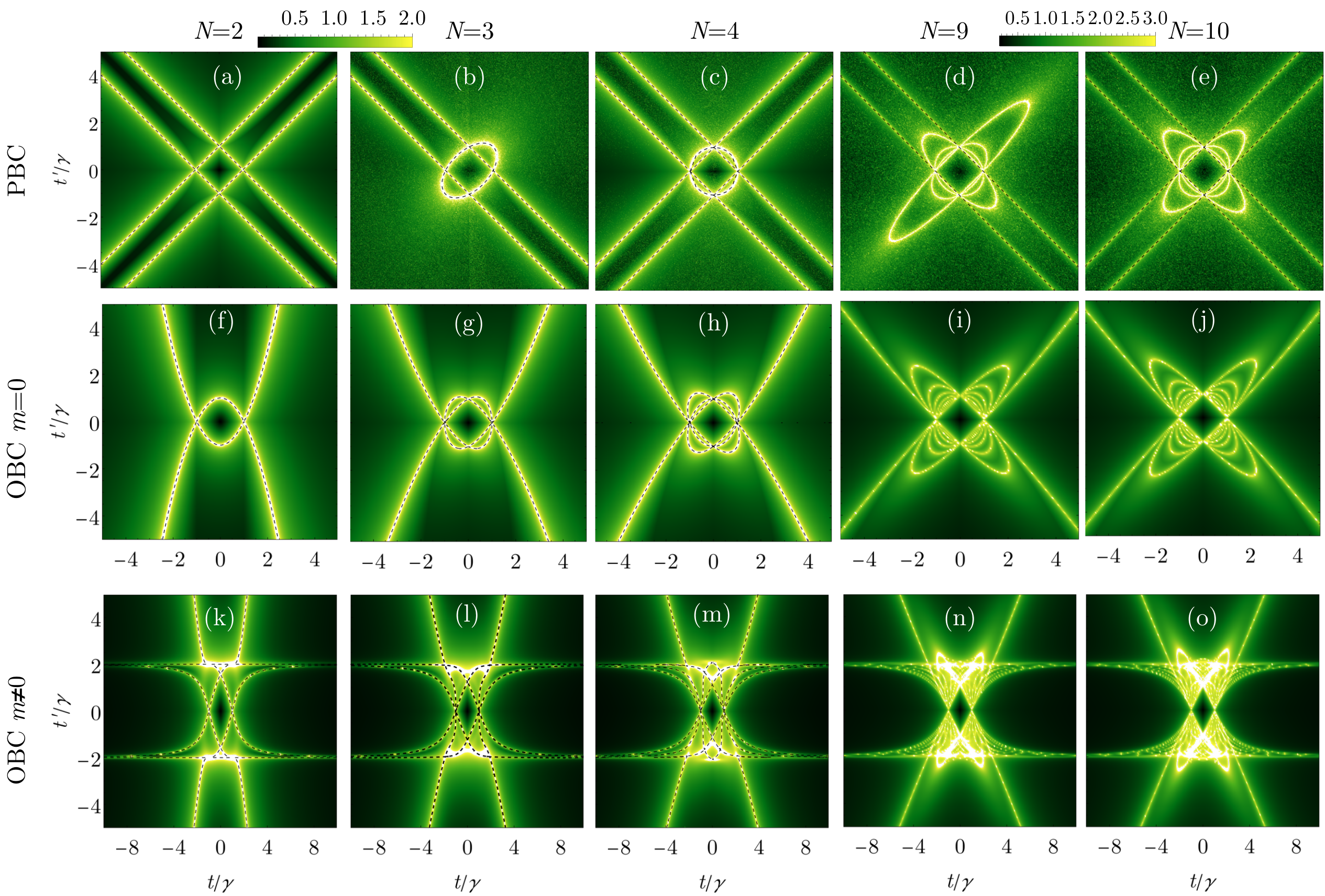}
  \caption{\textbf{Finite-size EPs under PBC and OBC.} 
  (a) For PBC, the natural logarithm of the condition number $\mathcal{K}(U) $ diverges at EPs. These divergences coincide with the analytic result from Eq.~\eqref{eq: EP's linear/ellipses constraint} (black dashed lines). For odd $N$, the lines with positive slopes $t\pm\gamma$ are absent. For $N>2$, the EPs form ellipses and rings. (a) $N=2$, (b) $N=3$, (c) $N=4$, (d) $N=9$, and (e) $N=10$. In all the panels $\epsilon_\pm=0$ and $m=0$. Note that these results are $m$-independent, since the value of $m$ only changes the degeneracy energy at the EPs. Note as well that for $N=9,10$, we indicate only the analytic lines corresponding to $\pm t\pm\gamma$. }
    \label{Fig3}
\end{figure*}
%
%
%
%
Generally, we focus on where to find EPs in the parameter space, particularly in the plane spanned by $t/\gamma$ and $t'/\gamma$. Setting $\epsilon_-=0$ in Eq.~\eqref{eq EP 1}, this becomes
\begin{align}\label{eq: EP's linear/ellipses constraint}
    1 = \frac{t^2}{\gamma^2} + \frac{(t')^2}{\gamma^2} + 2\frac{tt'}{\gamma^2} \cos(k)
\end{align}
and finite chain length and periodic boundary conditions impose $k = 2\pi n/N$, $n=1,\ldots,N$. For $k = \pi,\, 2\pi$, the right-hand side of Eq.~\eqref{eq: EP's linear/ellipses constraint} reduces to a complete square, and setting $y=t'/\gamma$, $x=t/\gamma$, the constraint can be simplified into four linear functions $y= s x + s'$ with $s,s' = \pm 1$ being independent. Otherwise, Eq.~\eqref{eq: EP's linear/ellipses constraint} describes an ellipse in parameter space. The rotation $u_\pm = (x\pm y)/\sqrt{2}$, corresponding to a $\pi/4$ rotation of the coordinate axes,  yields the ellipse equation
\begin{align}\label{eq: EP ellipse}
    1 = \frac{u_+^2}{\alpha_+^2} + \frac{u_-^2}{\alpha_-^2}, \quad \alpha_\pm  = \frac{1}{\sqrt{1\pm \cos(k)}}
\end{align}
with finite semi-axes $\alpha_\pm$ only if $k \neq \pi,\, 2\pi$.

To characterize the presence of EPs for finite-size systems with either PBC or OBC numerically, we investigate the condition number $\mathcal{K}(U) $.~\cite{schomerus2024eigenvalue,Wiersig_2023}  This is obtained by considering the eigenvector matrix $U$ of dimension $2N$, whose columns are the right eigenvectors of the Hamiltonian in Eq.~\eqref{Ham_TB}. The condition number is defined as the ratio of the largest to the smallest singular value of $U$, computed via the singular value decomposition: 
%
%
\begin{align}\label{Condition Number}
  \mathcal{K}(U) = \frac{s_\text{max}}{s_\text{min}}.  
\end{align}
%
%
This quantity provides a global measure of eigenstate 
non-orthogonality, related to the pre-eigenvalue condition number~\cite{Golub2013} that is equivalent to the inverse 
phase rigidity and the Petermann factor of quantum noise theory.~\cite{schomerus2024eigenvalue, Wiersig_2023} The natural logarithm $\ln\mathcal{K}(U)$ is close to zero when the Hamiltonian is diagonalizable and far from any EP; however, it diverges when approaching EPs, because the Hamiltonian is defective at an EP where the smallest singular value vanishes, allowing us to identify the EPs in parameter space. 
Subsequently, by investigating the Jordan decomposition of the Hamiltonian at the EP locations, we can determine their 
order and degeneracy.

In Fig.~\ref{Fig3}, we present numerical results via $\ln\mathcal{K}(U)$, 
revealing the EP structure for both PBC and OBC. We show that under PBC, the EPs 
lie on lines and ellipses, as predicted by Eq.~\eqref{eq: EP ellipse}. For increasing $N$, ellipses progressively fill the region between the lines, and 
the latter can be interpreted as the limiting case of an ellipse opening as 
$k\to\pi$ or $k\to 2\pi$. Specifically, $k\rightarrow \pi$ causes 
$\alpha_+\rightarrow \infty$, $\alpha_- \rightarrow 1/\sqrt{2}$, while 
$k\rightarrow 2\pi$ causes $\alpha_-\rightarrow \infty$, $\alpha_+\rightarrow 
1/\sqrt{2}$, such that $u_-^2=\alpha_-^{2}=1/2$ and $u_+^2=\alpha_+^{2}=1/2$, 
respectively, which corresponds to the equation of the lines in the rotated frame. 
Furthermore, all ellipses and lines pass through the points 
$(t/\gamma, t'/\gamma) = (0, \pm 1)$ and $(\pm 1, 0)$, marking parameter points 
where multiple EP$_2$ occur simultaneously. Lastly, as predicted analytically and 
confirmed numerically, the EP curves are independent of $m$ under PBC. The sole 
effect of $m$ is to shift the energy of the degeneracy point, in agreement with 
Eq.~\eqref{EP equation transition}. We also note an even-odd effect depending on 
the parity of $N$: for even $N$, the lines with both positive and negative slope 
host EPs, whereas for odd $N$, only the lines with negative slope do. This follows 
from the fact that $k = \pi$, finite size, and PBC together require $n=N/2$ to be 
an integer, and therefore explain the absence of the two lines on which EPs are 
found, as shown in Fig.~\ref{Fig3}(b).

\subsection{EPs under OBC and finite size}
\label{EP for OBC}
Here, we investigate EPs for finite-size systems under  OBC. We explicitly show that the nnn hopping term $m$ plays an active role, not only shifting the energy of the EPs but also generating new ones. For this purpose, we present in Figs.~\ref{Fig3}(f)--(j) results for $m=0$ and in Figs.~\ref{Fig3}(k)--(o) for $m\neq0$.

From Figs.~\ref{Fig3}(f)--(j), we observe drastic changes in the position of the EPs in the $(t,t')$--parameter space when switching from PBC to OBC for $m=0$. Only the degeneracy points at ($t/\gamma,t'/\gamma$)=\{($\pm1,0$),($0,\pm1$)\} remain unaffected. Most notably, for $N=2$, the four EP lines present in the PBC case transform into two mirror-symmetric parabolas. These parabolas host two degenerate EP$_2$. For $N>2$, these degeneracies are split at $(t/\gamma,t'/\gamma)=\{(0,\pm),(0,\pm)\}$ into multiple EP lines, and their shape depends on $N$; similarly, the asymptotic $t'/\gamma$ value of the parabolic branches decreases with increasing $N$, eventually becoming nearly coincident with the EP lines in the limit $N\to\infty$. Additionally, we observe how the EP lines at $t'/\gamma=0$ and $t/\gamma=\pm1$ evolve into ellipses similar to those identified in the PBC case.

The changes are even more drastic for $m\neq0$, see Figs.~\ref{Fig3}(k)--(o). 
One of the two outer branches evolves into quasi-horizontal lines with an asymptotic value $t'=2m$. The quantization condition for $k$ under OBC is generically transcendental and parameter-dependent, rendering analytical 
solutions intractable for an arbitrary system size $N$. This is 
not uncommon in related tight-binding systems: transcendental quantization conditions arise in the finite Kitaev chain~\cite{Leumer-2020} and in semiconducting Rashba nanowires,~\cite{Leumer2026} despite their different physical contexts. Crucially, the dependence of $k$ on $m$ through the 
dispersion relation $E_\pm(k) = \epsilon_+ - 2m\cos(k) + E_{\mathrm{RM},\pm}(k)$, cf.\ Eq.~\eqref{eq: RM dispersion relation}, is what distinguishes the OBC from the PBC case: if $k$ were independent of $m$, only the energy of the exceptional points could shift, but not their location in parameter space | which contradicts our findings. For this 
reason, we focus on numerical results throughout, with analytical results provided only for small $N$ (cf.\ black dashed lines).

The physical meaning of $t'=2m$ is revealed by analyzing the OBC spectrum of the model, \emph{e.g.}, as a function of $t$ at fixed $m$. For $t' \neq 2m$, the OBC spectrum is gapped for a certain range of $t$, whereas $t' = 2m$ closes it. This can be seen as signature of a topological phase transition driven by $m$ and indeed, isolated states can appear within the gap as we shall see below. 
%
%
\begin{figure*}[t]
    \centering
    \includegraphics[width=\linewidth]{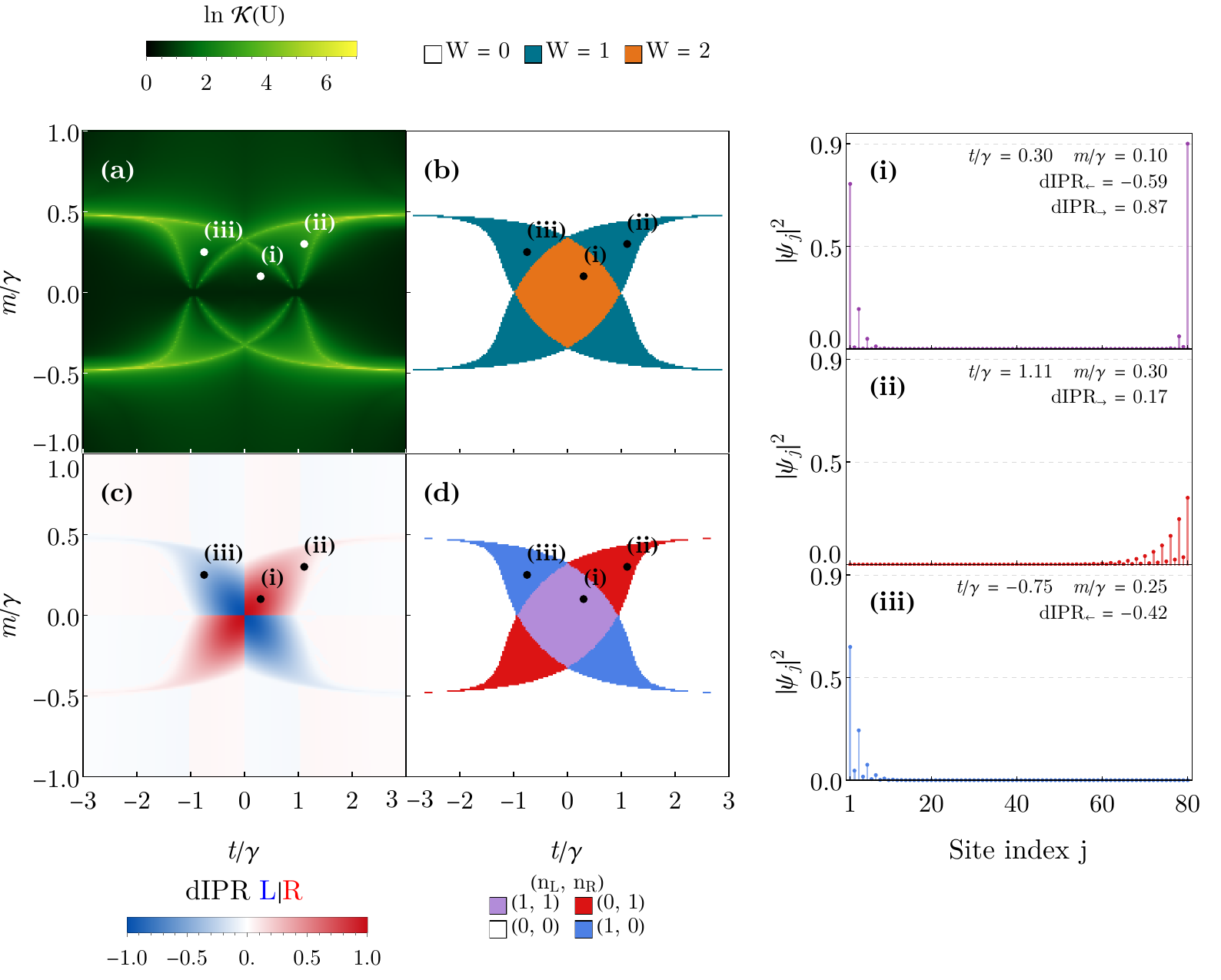}
    \caption{
    \textbf{Bulk winding number and its finite-size signatures: exceptional points and edge-state localization.}
    (a) Natural logarithm of the condition number $\mathcal{K}(U)$ of the
    right-eigenvector matrix $U$. (b) Bulk winding number $W$ computed from the characteristic polynomial~\cite{zhong2025topological}.
    (c) Directional inverse participation ratio $\mathrm{dIPR}(\delta)$,
    where blue (red) indicates preferential left- (right-)edge localization.
    (d) Left- and right-edge counts $(n_L,n_R)$, obtained by selecting
    states in $\mathcal{S}_f$ with $\mathrm{IPR}_j >
    f\,\overline{\mathrm{IPR}}$, with $f=2.5$, and classifying their
    directional signs $\sigma_j(\delta)$
    [Eqs.~\eqref{eq:edge_set}--\eqref{eq:nL_nR}].
    Purple, red, white, and blue, respectively, denote
    $(n_L,n_R)=(1,1)$, $(0,1)$, $(0,0)$, and $(1,0)$.
    The calculations use open boundary conditions with
    $N=40$ unit cells, $\epsilon_A/\gamma=0$,
    $\epsilon_B/\gamma=1.5$, $t'/\gamma=1$, and $\delta=0.5$.
    Panels (i)--(iii) show eigenvector probability densities
    $|\psi_j|^2$ at the marked points:
    (i) states localized at both boundaries,
    $(t/\gamma,m/\gamma)=(0.30,0.10)$;
    (ii) a right-localized state,
    $(1.11,0.30)$; and
    (iii) a left-localized state,
    $(-0.75,0.25)$. The number of sample points in both parameter sweeps is 200.}
    \label{fig:combined_winding}
\end{figure*}
\section{Emergence of edge states under OBC}
\label{sec: edge states}
To characterize the emergence of edge states under OBC, we combine four complementary diagnostics, and we present the results in Fig.~\ref{fig:combined_winding}. Figure~\ref{fig:combined_winding}(a) shows the location of the finite-size system's EP2s in parameter space, by providing the divergence of the natural logarithm of the condition number; as established previously, these EPs appear at the boundaries between topologically distinct phases even for the case with $\epsilon_-\neq 0$ due to the role of $m$ in the finite size system.~\cite{Kawabata:2019, Bessho:2020} In Fig.~\ref{fig:combined_winding}(b), the winding number | defined in the following section | characterizes the number of topological edge states directly from the bulk Hamiltonian. Figures.~\ref{fig:combined_winding}(c) and~\ref{fig:combined_winding}(d) then verify this bulk prediction with the actual finite-size spectrum:~\ref{fig:combined_winding}(c) reports the directional inverse participation ratio (dIPR) of the most localized eigenstate at each parameter point, whereas~\ref{fig:combined_winding}(d) gives the resulting left- and right-edge counts $(n_L,n_R)$ obtained from a threshold-based classification of the full OBC spectrum. 
\subsection{Winding number analysis and phase diagram}
Edge states in Hermitian 1D lattice models under OBC are a manifestation of a global bulk property, encoded in a nontrivial topological invariant.~\cite{zak1989berry,schnyder2008classification} For instance, a winding number or a Zak phase characterizes the global evolution of the bulk eigenstates when sweeping their momentum through the Brillouin zone (BZ).~\cite{zak1989berry,asboth2016short} These quantities provide an integer number that counts the encirclements around the reference point in parameter space, distinguishing topologically distinct bulk phases and, through the bulk--boundary correspondence, predicting the emergence of edge states in a finite system under OBC.~\cite{asboth2016short} The quantization of these invariants is strictly related to the presence of specific symmetries of the bulk Hamiltonian, such as chiral or particle-hole symmetry, which pin the edge state to a certain energy and provide protection against perturbations unless the bulk gap is globally closed.~\cite{altland1997nonstandard,schnyder2008classification,Kitaev2001}

When introducing nH terms, the reference point for the winding of bulk eigenstates in parameter space can change, requiring a modification of the traditional Zak phase by generalizing the momentum space. The GBZ re-defines the momentum quantum number $k$  to a complex value $k \rightarrow \kappa = k - \mathrm{i}\ln|r|$
to account for the possible macroscopic localization of bulk eigenstates under OBC.~\cite{yao2018edge,
bergholtz2021exceptional, zhong2025topological, okuma2020topological} Building on this framework, several nH topological invariants have been proposed.~\cite{yao2018edge, kunst2018biorthogonal, song2019non, kawabata2019symmetry} For example, the non-Bloch Zak phase is defined as a contour integral in the GBZ, using the geometric phase accumulated by the bulk eigenstate as $\kappa$ traverses the GBZ loop $\ee^{\mathrm{i}\kappa}$ in the complex plane. Notably, when the NHSE is absent, this loop reduces to the unit circle.~\cite{yao2018edge, okuma2020topological, song2019non} Despite the lack of symmetries, edge states can still exist, however, their energies are no longer pinned to a fixed value, rather they change as a function of parameters~\cite{zhong2025topological, bergholtz2021exceptional, kawabata2019symmetry} and, unlike the symmetry-protected invariants discussed above, their protection can no longer rest on chiral or particle-hole symmetry alone. Instead, a more general winding number has been proposed, defined directly from the analytic (Riemann-surface) structure of the characteristic equation rather than from any symmetry of the bulk Hamiltonian.~\cite{zhong2025topological} Crucially, this invariant does not replace but rather generalizes the conventional symmetry-protected classification: when the relevant symmetries are restored, it reduces to the familiar Hermitian and non-Hermitian symmetry-protected winding numbers, while remaining well defined even in their absence.~\cite{zhong2025topological, yao2018edge} In turn, the number of edge states under OBC can change even without the spectral gap closing.~\cite{zhong2025topological}

In our model, when all parameters are finite and $\epsilon_\text{A}\neq\epsilon_\text{B}$, the symmetries responsible for quantizing the conventional symmetry-protected topological invariants are absent (see Table~\ref{tab:symmetries}). The topology can instead be characterized using more general geometric constructions. Accordingly, in the present work, we adopt the Riemann-surface-based invariant introduced by Zhong~\emph{et al.} in Ref.~[\onlinecite{zhong2025topological}], which remains applicable without relying on these symmetries.

For each band $j=\pm$, let $\boldsymbol{\psi}_{j}(k)=\left(\psi_{j,\mathrm{A}}(k),\psi_{j,\mathrm{B}}(k)\right)^\mathrm{T}$ denote the corresponding bulk right eigenvector. Since an overall rescaling of $\boldsymbol{\psi}_{j}(k)$ leaves the physical state unchanged, only the ratio of its components carries physical information. We therefore define the \emph{upper-to-lower} eigenvector component ratio for mode $j$:
\begin{equation}
M_j\equiv\frac{\psi_{j,\mathrm{A}}}{\psi_{j,\mathrm{B}}}\equiv M_j(\ee^{\ii k}),
\end{equation}
which sets a closed curve $M_j(\ee^{\ii k})$ when $k$ sweeps $[0,2\pi)$. We employ the following Riemann-surface winding number
\begin{equation}
W_j=\bigl(1+W_{\mathrm{deg},j}-W_{\mathrm{branch},j}\bigr)\bmod2,
\label{eq:Wjband}
\end{equation}
where $W_{\mathrm{deg},j}$ and $W_{\mathrm{branch},j}$ denote the net encirclement numbers of $M_j(\ee^{\ii k})$ around the eigenvector-degeneracy points $M_{\mathrm{deg}}^{(n)}$, $n\in\mathcal D_j$, and the branch points $M_{\mathrm{br}}^{(n)}$, $n\in\mathcal D_j$, respectively, computed as
5
%
\begin{align}
W_{\beta,j} &= \frac{1}{2\pi} \sum_{n\in\mathcal D_j} \oint \frac{d}{dk} \arg\!\left( M_j(\ee^{\ii k})-M_{\beta}^{(n)} \right)dk,
\label{eq:WdegWbr}
\end{align}
%
%
with $\beta=\{\mathrm{deg,br}\}$ and $\mathcal D_j$ the index set of degeneracy/branch points relevant to sheet $j$, such that $\{M_{\mathrm{br}}^{(n)}\}_{n}$ and $\{M_{\mathrm{deg}}^{(n)}\}_{n}$ are, respectively, the full sets of branch points and eigenvector-degeneracy points of the energy-free polynomial $g(M,z)=0$, obtained by eliminating the energy from the eigenvalue equation $H(z)\boldsymbol{\psi}=E\boldsymbol{\psi}$. Viewed as an algebraic curve, $g(M,z)=0$ defines a two-sheeted Riemann surface whose topology determines the invariant. Since the NHSE is absent in our reciprocal model, the curves $M_j(\ee^{\ii k})$ are traced directly on the ordinary BZ rather than on the GBZ.

For the present model with nnn hopping, the two analytic branches of the eigenvector ratio are
%
%
\begin{equation}
M_{\pm}(\ee^{\ii k})= \frac{-\Gamma\mp \sqrt{\Delta(\ee^{\ii k})}}{t+t'\ee^{-\ii k}},
\label{eq:McurvesMain}
\end{equation}
%
%
where
%
%
\begin{equation}
\Delta(z)=\Gamma^{\,2}+\left(t+t'z^{-1}\right)\left(t+t'z\right).
\label{eq:DeltaMain}
\end{equation}
%
%
These closed curves define the integration paths entering the winding-number definition in Eq.~\eqref{eq:WdegWbr} (shown explicitly on the Riemann sphere in Fig.~\ref{fig:Mriemann_appendix}).
The associated characteristic equation reduces to a four-root polynomial in $z$, rather than the generic eight-root case. The corresponding branch points are
%
%
\begin{equation}
M_{\mathrm{br}}^{(s,\eta)}= \frac{-\delta_s+\eta\sqrt{\delta_s^2+4t^2}}{2t},
\qquad
s,\eta\in\{\pm1\},
\label{eq:Mbrmain}
\end{equation}
%
%
where $\delta_s=2(\Gamma+s\,\ii t')$.
The indexing represented in the Riemann sphere of Fig.~\ref{fig:Mriemann_appendix} follows $n=1,\dots,4$ with signs
$(s,\eta)=(+1,+1)$ for $n=1$, $(+1,-1)$ for $n=2$, $(-1,+1)$ for $n=3$, $(-1,-1)$ for $n=4$.
These four solutions arise from the factorization of the branch-point condition $\mathrm{Disc}_z[g]=0$. 

The eigenvector-degeneracy points are
%
%
\begin{equation}
M_{\mathrm{deg}}^{(\sigma,\eta)} = \frac{\sigma t'+\eta\sqrt{t'^2-4m^2}}{2m},
\qquad
\sigma,\eta\in\{\pm1\},
\label{eq:Mdegmain}
\end{equation}
%
%
The sign $\sigma$ selects one of the two quadratic factors in $[m(M^2+1)-t'M][m(M^2+1)+t'M]=0$ (see Eq.~\ref{eq:quadFacs} in Appendix~\ref{app: winding number}), while $\eta$ selects the root within that factor. Representing through the Riemann spheres in Fig.~\ref{fig:Mriemann_appendix} we consider the same indexing convention for $(\sigma,\eta)$ as in $(s,\eta)$ for the branch points.
The degeneracy points arise from the condition that two distinct complex momenta $z$ correspond to the same eigenvector ratio $M$ on the algebraic curve $g(M,z)=0$, characterizing degeneracies of the eigenvector mapping rather than exceptional points associated with band coalescence. 

For the evaluation of the winding number, the correspondence between the singular points and the analytic branches $M_\pm(\mathrm e^{\mathrm i k})$ is established by continuously tracking the two $M$-curves along the Brillouin zone and employing the same analytic continuation convention used to define them. In particular, the algebraic indexing of $M_{\mathrm{br}}^{(n)}$ and $M_{\mathrm{deg}}^{(n)}$ does not determine their sheet assignment, this is fixed only after analytic continuation, which is precisely how the index set $\mathcal D_j$ in Eq.~\eqref{eq:WdegWbr} is determined. The derivation of the algebraic curve $g(M,z)$, the analytical expressions for $M_{\mathrm{br}}$ and $M_{\mathrm{deg}}$ are presented in Appendix~\ref{app: winding number}. Figure~\ref{fig:Mriemann_appendix} illustrates this construction for the three representative parameter points shown in Fig.~\ref{fig:combined_winding}(i)--(iii).

In general, the topological phase diagram is determined by how the trajectory $M_j(\ee^{\ii k})$ winds around the branch and eigenvector-degeneracy poles. As shown in Eq.~\eqref{eq:Mbrmain}, the branch points depend on the intra-cell hopping $t$, the inter-cell hopping $t'$, and the non-Hermitian on-site term $\Gamma=\epsilon_-+\ii\gamma$, while the eigenvector-degeneracy points in Eq.~\eqref{eq:Mdegmain} depend only on the competition between the nnn hopping $m$ and the inter-cell hopping $t'$. Therefore, the winding number is governed by the interplay between these two sets of microscopic parameters. As they vary, the branch and eigenvector-degeneracy poles move in the $M$ plane relative to the closed trajectories $M_j(\ee^{\ii k})$, modifying their encirclement numbers and hence the topological invariant. A non-trivial winding, $W_j\neq0$, signals a topological edge state associated with band $j=\pm$, while the total number of edge states is
$W_{\mathrm{total}}=W_+ + W_-$.

While in the NH SSH limit ($\epsilon_\pm=m=0$), the hopping ratio $t/\gamma$ alone determines the topology, restricting the system to only two sectors, $W=2$ around $t/\gamma=0$, where the chain supports one edge state at each boundary, and $W=0$ once the boundary sites are sufficiently coupled to the bulk. Introducing a finite nnn hopping $m$ and staggered potential $\epsilon_-$ broadens this picture, giving the possibility for an intermediate sector to arise with $W=1$ in which only one topological edge state is present. In Fig.~\ref{fig:combined_winding}(b), we show the winding number (renaming $W_{\mathrm{total}}=W$ in the following) as a function of $t/\gamma$ \emph{vs.} $m/\gamma$ for $\epsilon_-\neq 0$.

The origin of this intermediate phase with $W=1$ is tied to the branch poles $M_{\mathrm{br}}^{(n)}$ rather than to the degeneracy poles $M_{\mathrm{deg}}^{(n)}$. From Eq.~\eqref{eq:Mbrmain}, the branch poles depend on $t$, $t'$, and $\Gamma$, while the degeneracy poles in Eq.~\eqref{eq:Mdegmain} depend only on $m$ and $t'$: the two pole families respond to different subsets of microscopic parameters. At $\epsilon_-=0$, the model recovers the spatial-inversion symmetry $\mathcal{I}=\sigma_x$ of the Rice--Mele part, exchanging sublattices A and B. Consequently, the winding number is then restricted to $W\in\{0,2\}$ over the full parameter sweep, and the intermediate $W=1$ sector only opens once $\epsilon_-\neq0$ breaks this symmetry. The nnn hopping $m$, by contrast, enters $\mathcal{H}(k)$ only as an identity shift and therefore commutes with $\mathcal{I}$: it relocates the degeneracy poles $M_{\mathrm{deg}}^{(n)}$ through the merger condition $t'^2=4m^2$ from Eq.~\eqref{eq:Mdegmain} and can drive $W$ between $0$ and $2$, but does not by itself produce $W=1$. Thus $m$ decides where the $W=2\to0$ boundary sits, while $\epsilon_-$ decides whether an intermediate $W=1$ plateau exists between them at all.
More generally, the winding number changes only when a pole crosses the closed trajectory $M_j(\ee^{\ii k})$, \emph{i.e.}, when either $M_{\mathrm{deg}}^{(n)}$ or $M_{\mathrm{br}}^{(n)}$ changes its encirclement number with respect to $M_j(\ee^{\ii k})$. For the degeneracy poles, this crossing coincides with their pairwise merger at $t'^2=4m^2$ ($t'=2m$ for positive hoppings), which is precisely the condition for the OBC bulk gap closing (Sec.~\ref{EP for OBC}); for the branch poles, the analogous merger condition governs the boundary of the $\epsilon_-$-induced $W=1$ window. Both mergers are confirmed directly by the condition-number map in Fig.~\ref{fig:combined_winding}(a): its divergent lines mark second-order exceptional points (EP2s), where the eigenvalues become degenerate and the eigenvectors coalesce, and these EP2 lines trace out exactly the phase boundaries of the winding-number diagram in Fig.~\ref{fig:combined_winding}(b).
This bulk picture is verified against the finite-size system diagnostics in Figs.~\ref{fig:combined_winding}(c) and~\ref{fig:combined_winding}(d), where the maximum value of the dIPR and the edge count independently confirm the number and boundary localization of edge states predicted by the winding number across all three regions.
\subsection{Directional inverse participation ratio and edge-counts}
Having established the bulk topological phase diagram through the $M$-Riemann winding number, we now verify the corresponding finite-size edge-state behavior under OBC. While the winding number predicts the number of topological edge states from the bulk Hamiltonian alone, the eigenstates of a finite chain allow us to determine if the edge states are present, on which boundary they localize, and how strongly they are confined. For this reason, we introduce the directional inverse participation ratio (dIPR), which combines the degree and the direction of localization into a single quantity. For a right eigenstate $|\psi_j\rangle$ of the finite OBC Hamiltonian with components $\psi_j(s)$ ($s=1,\dots,2N$), the inverse participation ratio (IPR) is
%
%
\begin{align}
    \mathrm{IPR}_j = \sum_{s=1}^{2N} p_j^2(s),\qquad
    p_j(s) = \frac{|\psi_j(s)|^2}{\sum_{s'}|\psi_j(s')|^2},
    \label{eq:IPR}
\end{align}
%
%
which approaches $1/(2N)$ for extended states and $1$ for perfectly localized states. To distinguish left from right localization, we set
$\mu_j(\delta) = \sum_{s=1}^{2N} \bigl(s-N-\delta\bigr)\,p_j(s)$ with a  small offset $\delta=1/2$ to avoid ambiguity for states exactly
centered. The directional sign $\sigma_j(\delta)=\operatorname{Sign}[\mu_j(\delta)]$ defines the directional IPR through $\mathrm{dIPR}_j(\delta) = \sigma_j(\delta)\,\mathrm{IPR}_j$. Thus $\mathrm{dIPR}_j<0$ identifies a left-localized state, $\mathrm{dIPR}_j>0$
a right-localized one. In Fig.~\ref{fig:combined_winding}(c), we show the highest-magnitude signed dIPR ($\mathrm{max}|\mathrm{dIPR}|$) obtained at each point of the parameter plane, saturating toward $\pm 1$ in regions of strong single-boundary localization and vanishing where the least localized (most extended) states dominate the spectrum.

Since every eigenstate of a finite chain has a finite participation ratio, even perfectly extended bulk states satisfy $\mathrm{IPR}\sim1/(2N)$ rather than exactly zero. Consequently, a finite threshold is required to distinguish true edge-localized states from the bulk ones. Rather than using an absolute criterion, we define the threshold relative to the bulk spectrum, which changes at each parameter point during the sweep. To isolate the IPR values of edge states from those of the bulk, we use the median IPR of the spectrum ($\overline{\mathrm{IPR}}$), since for a chain of $N=40$ the median is dominated by bulk IPR values. Multiplying by a factor $f=2.5$ then sets a threshold above which only edge-state IPR values remain. The edge index set is
%
%
\begin{equation}
  \mathcal{S}_f \;=\; \bigl\{\, j \;:\; \mathrm{IPR}_j > f\,\overline{\mathrm{IPR}} \,\bigr\},
  \label{eq:edge_set}
\end{equation}
%
%
which we take as the operational definition of an edge state, independent of the winding number $W_j$ shown in
Fig.~\ref{fig:combined_winding}(b). Each state $j\in\mathcal{S}_f$ is classified by its directional sign $\sigma_j(\delta)$, yielding the
left- and right-edge counts
%
%
\begin{align}
  n_\text{L} \;=\; \bigl|\{\, j\in\mathcal{S}_f : \sigma_j(\delta)=-1 \,\}\bigr|,
  \nonumber \\ \qquad
  n_\text{R} \;=\; \bigl|\{\, j\in\mathcal{S}_f : \sigma_j(\delta)=+1 \,\}\bigr|,
  \label{eq:nL_nR}
\end{align}
%
%
the finite-chain counterpart of the bulk winding numbers $(W_j^{\mathrm{left}}, W_j^{\mathrm{right}})$.

While $\max|\mathrm{dIPR}|$ in Fig.~\ref{fig:combined_winding}(c) flags where and on which boundary the most localized eigenstate sits, it is a single scalar per parameter point and therefore cannot resolve how many edge states are present: where the winding number predicts two edge modes, it reports only the sign and strength of the more sharply confined one, discarding the second. Counting edge states therefore requires the full distribution of $\mathrm{IPR}_j$ rather than its maximum alone, motivating the threshold-based classification of Fig.~\ref{fig:combined_winding}(d). Figure~\ref{fig:combined_winding}(c) remains complementary, however, since it tracks the sign of the most localized state in a continuous way across the parameter sweep and so visualizes its dependency and boundary crossovers that the $(n_\text{L},n_\text{R})$ count in Fig.~\ref{fig:combined_winding}(d) misses.

Comparing Figs.~\ref{fig:combined_winding}(d) to~\ref{fig:combined_winding}(b) confirms the bulk-boundary correspondence across all three sectors: $W=2$ yields one left- and one right-localized edge state, $W=0$ yields none, and the intermediate $W=1$ region yields a single localized edge state. Thus, while the winding number counts the topological edge modes from the evolution of the $M$-Riemann trajectories, the dIPR identifies the corresponding eigenstates in the finite spectrum and pinpoints the boundary on which each localizes.

\subsection{OBC vs PBC transition spectra}

\begin{figure}[t]
    \centering
    \includegraphics[width=\linewidth]{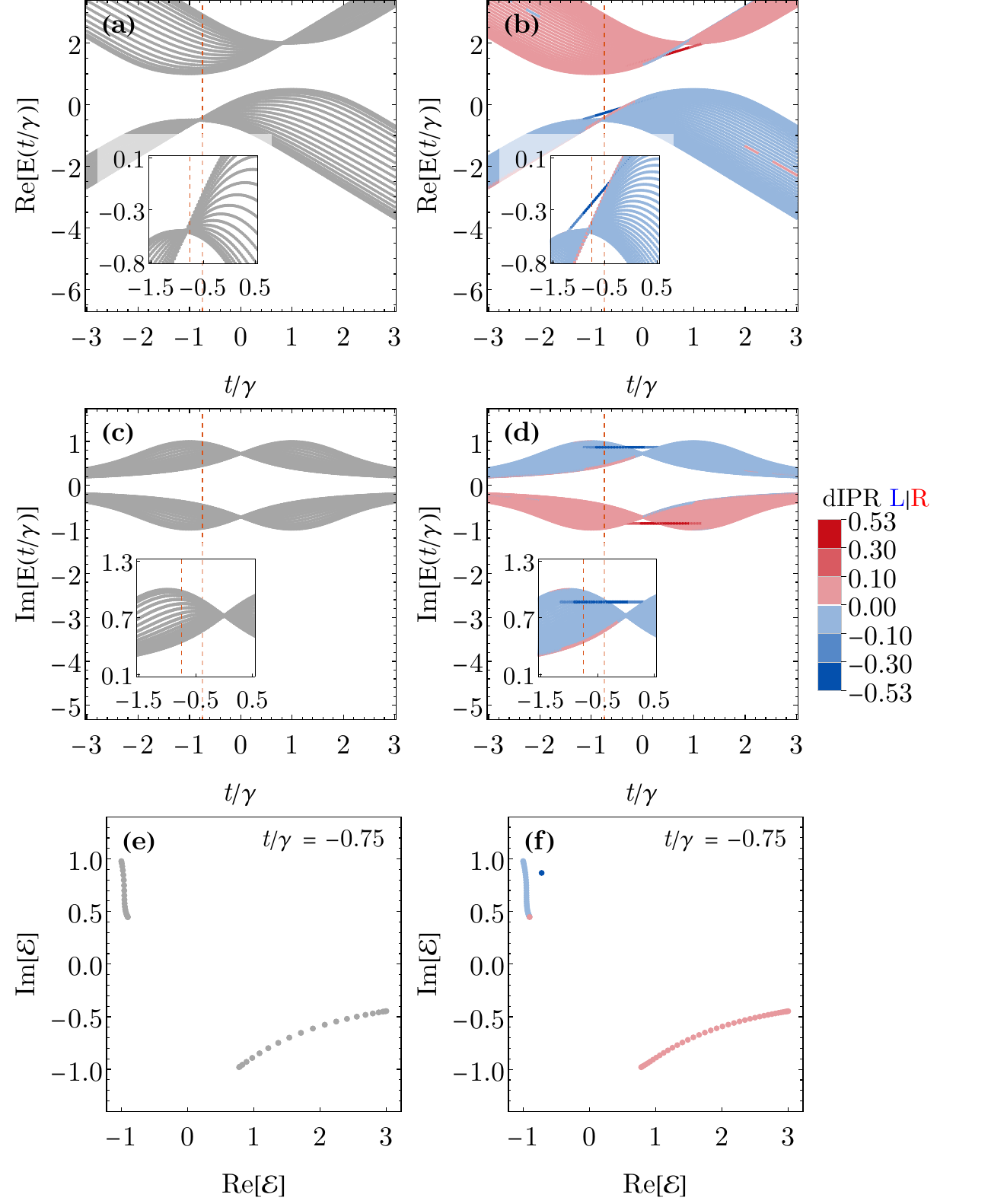}
    \caption{
   \textbf{Topological phase transition spectra and edge state emergence.}  Spectra of the non-Hermitian next-nearest-neighbour Rice--Mele model for a chain of \(N=40\) unit cells. The parameters are \(\epsilon_A=0\), \(\epsilon_B=1.5\), \(\gamma=1\), \(m=0.25\), and \(t'=1\). The hopping amplitude \(t\) is swept using 200 parameter values. Since \(t'=\gamma=1\), this is numerically equivalent to the displayed range \(-3\leq t/t'\leq3\). The left column displays the PBC spectrum and the right column the corresponding OBC spectrum. The top and middle rows show \(\mathrm{Re}[E/\gamma]\) and \(\mathrm{Im}[E/\gamma]\), respectively, as functions of \(t/\gamma\).The insets in each panel provide a magnified view of the phase-transition region, where the edge states merge into the bulk spectrum. The bottom row shows the complex-energy spectrum at \(t/\gamma=t/t'=-0.75\), indicated by the vertical dashed line in the upper panels.}

    \label{fig:obc_pbc_phase_transition_dIPR}
\end{figure}

Figure~\ref{fig:obc_pbc_phase_transition_dIPR} shows the evolution of the OBC and PBC spectra as the hopping amplitude $t$ is varied, while all remaining parameters are held fixed. The OBC spectrum contains two branches with a large signed directional inverse participation ratio (dIPR), whereas the PBC spectrum contains only the bulk spectral continuum. The opposite signs of the dIPR identify the two boundary modes as states localized at opposite ends of the chain.
%
%
%
%

Comparing OBC and PBC spectra (at fixed parameters) in the complex energy plane (bottom row of Fig.~\ref{fig:obc_pbc_phase_transition_dIPR}) corresponding to the parameter point indicated by the vertical dashed line in the upper panels, one observes that both spectra form identical arcs in the complex plane. This snapshot coincides with point~(iii) in Fig.~\ref{fig:combined_winding}, \((t/\gamma,m/\gamma)=(-0.75,0.25)\), where the representative left-localized edge-state profile is shown.
The equivalence of the two spectra indicates the absence of the NHSE and the coincidence of the GBZ with the ordinary BZ; nevertheless, one isolated eigenvalue appears exclusively in the OBC spectrum, carrying large positive dIPR, which confirms the boundary origin and rules out any accidental bulk degneracy. The dIPR contrast between edge and bulk eigenstates | with \textbf{$|\mathrm{dIPR}_\mathrm{edge}| \approx 1/3,|\mathrm{dIPR}_\mathrm{bulk}|^{-1}$} relative to the extended states | provides a direct finite-size diagnostic: well-localized edge states saturate the IPR bound $\sim 1/(2N)^{-1}$, while bulk eigenstates remain delocalized across the $N = 40$ unit-cell chain.

The evolution of $\mathrm{Re}[E]$ and $\mathrm{Im}[E]$ as a function of $t/\gamma$ (top and middle rows) further illustrates the non-trivial topology: as $t/\gamma$ crosses the phase boundary (see insets of the OBC spectra in Fig.~\ref{fig:obc_pbc_phase_transition_dIPR}), the edge state eigenvalues emerge discontinuously from the bulk continuum, consistent with the winding-number jump $\Delta W = 1$ computed via Eq.~\eqref{eq:Wjband}. Outside the topological phase ($t/\gamma > t_c/\gamma$), all eigenvalues are bulk-like (faded, small $|\mathrm{dIPR}|$) and the spectrum is indistinguishable between OBC and PBC at the scale of the figure, confirming the bulk-boundary correspondence established by the non-Bloch winding number framework.~\cite{yao2018edge, zhong2025topological}

\section{Discussion and Conclusion}
\label{sec: conclusion}
In this paper, we have studied the exceptional points of a generalized non-Hermitian Rice-Mele model with a next-nearest-neighbor hopping term $m$. We have presented analytic results for a bulk system and for both periodic and open boundary conditions. The intricacy of parameter dependencies --- particularly in $m$ --- under open boundary conditions demands generally a numerical treatment, at least for sufficiently large system sizes~$N$. Under OBC, we identified the relation $t'=2m$ between the presence of additional EPs in the OBC parameter space and a gap closing in the OBC spectrum, which coincides with the $M_\textrm{br}$ points merging of the reflecting the generally non-trivial topology of our model. The scheme of extended symmetry classes for nH systems~\cite{kawabata2019symmetry} revealed the general lack of symmetries, with the time reversal symmetry TRS$^\dagger$ being the only exception. Conventionally a contradiction to the presence of topological edge states in Hermitian systems, the framework presented in Ref.~[\onlinecite{zhong2025topological}] defines a topological phase diagram with both trivial and non-trivial phases on the basis of a winding number analysis for the bulk Hamiltonian. We have provided evidence that the conventional bulk-boundary correspondence is intact, \emph{i.e.}, topological edge states emerge under OBC for parameters assigned to a non-trivial winding number $W=1,2$, using (i) the inverse participation ratio to quantify the localization of states and (ii) a comparison of PBC and OBC energy spectra.

A particular result is that | for OBC and PBC | EPs are generally of second order only. Yet, the rank of the Hamiltonian can be sufficiently reduced as several EPs may occur simultaneously, depending on the choice of parameters. While rather restricted and independent of the nnn hopping $m$ under PBC, conditions for EPs and points of multiple EPs in the $t'/\gamma$-$t/\gamma$-plane can be tuned by sweeping through values of $m$ under OBC. Conventionally, a variation of $m$ may be difficult to realize, since, in atomic systems, changing the interatomic distance alters both nearest- and next-nearest-neighbor hopping terms in an uncontrolled manner. However, it has been shown in the past that effective nnn hopping amplitudes can be engineered~\cite{Leumer2023} based on (real) onsite energies and nearest-neighbor-hopping terms in quantum ladder setups that appear similar but not identical to our system. There, harnessing gate potentials allows to precisely control $m=m(\epsilon_\pm)$. In turn, the control of $m$ allows for re-localizing EPs in parameter space for OBC to exploit the improved sensing of EPs as small perturbations provide a magnified response.~\cite{Grom2025, wiersig2014, Liu2016}
\section{Acknowledgments}
The work of DB and CMS is supported by the Grant PID2024-162933NB-I00 (QUILL) funded by MICIU/AEI/10.13039/501100011033 by ERDF/EU. DB acknowledge the financial support of the Basque Government’s Department of Education through project PIBA\_2023\_1\_0007 (STRAINER), the IKUR Strategy under the collaboration agreement between the Ikerbasque Foundation and DIPC, on behalf of the Basque Government’s Department of Education and the Gipuzkoa Provincial Council, within the QUAN-000021-01 project; and the ``Artificial Quantum Matter: From 2D Materials to Spin Lattice Systems" - BBVA Foundation Fundamentos Program 2024". NL acknowledges fruitful discussions with M. Marga\'nska and M. Mierzejewski.

\appendix
\makeatletter
\renewcommand{\thefigure}{\Alph{section}\arabic{figure}}
\renewcommand{\theequation}{\Alph{section}\arabic{equation}}
\renewcommand{\thetable}{\Alph{section}\arabic{table}}
\@addtoreset{figure}{section}
\@addtoreset{equation}{section}
\@addtoreset{table}{section}
\makeatother
\section{Coalescence of bulk eigenstates}
\label{app: Coalescence of bulk eigenstates}
For generic parameters, eigenvectors of the bulk Hamiltonian $\mathcal{H}(k)=[ \epsilon_+ -2m\cos(k) ]\mathbbm{1}_2 + \mathcal{H}_{\rm RM}$ are those of $\mathcal{H}_{\rm RM}$, only at shifted energy $E_{\rm RM,\pm}\rightarrow  E_{\rm RM,\pm} + \epsilon_+ -2m\cos(k)$. Straightforwardly one finds, the right eigenvectors
\begin{align*}
 \psi_+ =\frac{1}{N_+}\begin{pmatrix}
     1\\ \frac{\epsilon_- +\ii \gamma -E_{\rm RM,+} }{t+t' 
     \exp{-\ii k}}
 \end{pmatrix},~\psi_- =\frac{1}{N_-}\begin{pmatrix}
      \frac{t+t' 
     \exp{-\ii k}}{\epsilon_- +\ii \gamma -E_{\rm RM,-} }\\1
 \end{pmatrix}.
\end{align*}
with normalization factors $N_\pm$ and for arbitrary parameters. At $E_{\rm RM,\pm}=0$, we have
\begin{align}
    \psi_- &\propto\begin{pmatrix}
      \frac{t+t' 
     \exp{-\ii k}}{\epsilon_- +\ii \gamma }\\1
 \end{pmatrix}
 =\frac{\epsilon_- +\ii \gamma }{t+t' 
     \exp{-\ii k}}\begin{pmatrix}
      1\\\frac{t+t' 
     \exp{-\ii k}}{\epsilon_- +\ii \gamma }
 \end{pmatrix}
\end{align}
and comparing $\psi_\pm$, we observe that both became linearly dependent, i.e., they coalesce; thus, marking $E_{\rm RM,\pm}=0$ as EP in the bulk.
\section{Derivation of the winding number}
\label{app: winding number}
Because the GBZ coincides with the unit
circle for the present model, the topological invariant can be
computed as a winding number on the ordinary BZ. For each band $j=\pm$, the closed curve $M_j(e^{ik})$ is traced as $k$ sweeps $[0,2\pi)$; its winding around the fixed
degeneracy and branch points defines the invariant in Eq.~\eqref{eq:Wjband}. In the following, we verify first that GBZ, BZ are identically using the palindromic form of the characteristic polynomial. We then derive the energy-free algebraic curve $g(M,z)$ and obtain the explicit expressions for the $M$-curves, the branch points $M_{\mathrm{br}}^{(n)}$, and the eigenvector-degeneracy points $M_{\mathrm{deg}}^{(n)}$ entering the winding-number construction.
\subsection{Absence of NHSE: coincidence of GBZ, BZ}
To uncover the edge states present in this model, we start by setting
$z=\ee^{\text{i}k}$, and compute the characteristic polynomial through
$\det[E\mathbbm{1}_2 - \mathcal{H}(z)]=0$. We multiply by $z^2$ to get rid of
the negative powers and obtain an explicit \emph{palindromic quartic}
%
%
\begin{equation}
P(z) = m^2 z^4 + a_1 z^3 + a_2 z^2 + a_1 z + m^2 = 0,
\label{eq:palindrome}
\end{equation}
%
%
with energy-dependent coefficients
%
%
\begin{align}
a_1(E) &= 2m(E-\epsilon_+) - tt',
\label{eq:a1_OBC}\\[4pt]
a_2(E) &= (E-\epsilon_+)^2 - |\Gamma|^2
- (t^2+t'^2) + 2m^2,
\label{eq:a2_OBC}
\end{align}
%
%
where we have used again that $2\epsilon_\pm = \epsA \pm \epsB$ and
$\Gamma = \epsilon_- + \mathrm{i}\gamma$.
The palindromic structure
$a_0=a_4=m^2$, $a_1=a_3$ follows directly from TRS$^\dagger$, which enforces $P(z)=z^4P(z^{-1})$ for all parameter
values.~\cite{yao2018edge, zhong2025topological} From  Vieta's theorem
we obtain
%
%
\begin{equation}
z_1 z_4 = 1, \qquad z_2 z_3 = 1,
\label{eq:vieta_OBC}
\end{equation}
%
%
for the four complex roots sorted by modulus
$|z_1|\leq|z_2|\leq|z_3|\leq|z_4|$. Physically, this means that every
propagating mode is paired with a decaying one of equal and opposite
amplitude, which prevents bulk states from accumulating at a single
boundary. To find the contour of the GBZ at a given
energy in the OBC, we set $|z_2|=|z_3|$, which together with the Vieta
condition ($z_2 z_3=1)$ forces $|z_2|=|z_3|=1$. These conditions prove that the GBZ will coincide with the standard BZ and that no NHSE will be present. Although the NHSE is absent, solutions with $|z|\neq1$ still determine the edge-state wavefunctions under OBC. The reciprocal pairing of the complex Bloch roots therefore remains essential for identifying the admissible decaying solutions entering the boundary problem, in direct analogy with the non-Bloch wave-factor construction developed in Refs.~[\onlinecite{yao2018edge,yokomizo2020non,yang2024anatomy}]. Since the GBZ coincides with the ordinary BZ, however, the topological invariant can be evaluated directly by tracing the curves $M_j(\ee^{\ii k})$ on the ordinary BZ.

\subsection{$M$-Riemann algebra formalism: $M_j$, $M_{\mathrm{deg}}^{(n)}$, $M_{\mathrm{br}}^{(n)}$}
The $M$-Riemann algebra formalism introduced in Ref.~[\onlinecite{zhong2025topological}] encodes the topology in an energy-independent algebraic relation between the eigenvector component ratio and the complex momentum.
Starting from the bulk eigenvalue equation
\begin{equation}
H(z)\boldsymbol{\psi}_j(z)=E\boldsymbol{\psi}_j(z),
\end{equation}
the two rows of the eigenvalue problem provide two expressions for the
eigenvector ratio $M_j$ introduced in Sec.~\ref{sec: edge states}, which
coincide whenever the bulk dispersion relation is satisfied:
\begin{align}
M(z) =
-\frac{t+t'z^{-1}}
{E-\epsilon_A-\mathrm{i}\gamma
+m(z+z^{-1})}
\label{eq:Mi_row1}
\\[3ex]
M(z)
=
-\frac{E-\epsilon_B+\mathrm{i}\gamma
+m(z+z^{-1})}
{t+t'z}
\label{eq:Mi_row2}
\end{align}

which are equivalent whenever the bulk dispersion relation is satisfied.

Eliminating the energy $E$ between
Eqs.~\eqref{eq:Mi_row1} and~\eqref{eq:Mi_row2}
yields an energy-independent polynomial relating the eigenvector ratio
and the complex momentum,
%
%
%
\begin{align}
g(M,z)=
t' z^2 M^2
+\left(tM^2+2 \Gamma M-t\right)z
-t'
\label{eq:gMz}
\end{align}

%
%
where
\(
\Gamma
=
\epsilon_-+\mathrm{i}\gamma.
\)

Equation~\eqref{eq:gMz} defines the algebraic curve underlying the
$M$-Riemann construction. Since the energy has been eliminated
explicitly, its topology depends solely on the analytic structure of
$g(M,z)$ in the $(M,z)$ plane. The winding invariant introduced in
Sec.~\ref{sec: edge states} is therefore determined entirely by the
properties of this algebraic curve. In the following, we derive its two
analytic branches together with the associated branch and
eigenvector-degeneracy points entering the winding-number
construction.

Since the energy-free polynomial $g(M,z)$ is quadratic in the eigenvector ratio $M$ for each fixed complex momentum $z$, solving for $M$ gives the two analytic branches of the Riemann sheets introduced in Eq.~\eqref{eq:McurvesMain},
\begin{equation}
M_{\pm}(z)
=
\frac{-\Gamma
\mp
\sqrt{\Delta(z)}}
{t+t'z},
\label{eq:Mcurves}
\end{equation}
where
\begin{equation}
\Delta(z)
=
\Gamma^{\,2}
+
\left(t+t'z^{-1}\right)
\left(t+t'z\right).
\label{eq:dz_Delta}
\end{equation}
Since the topology is evaluated on the ordinary BZ, $z=\ee^{\ii k}$ with
$k\in[0,2\pi)$, generates the two closed curves
$M_\pm(\ee^{\ii k})$ entering the winding-number construction.

The branch points of the Riemann surface correspond to the values of $M$ at which the two sheets merge. Its discriminant with respect to $z$, therefore, identifies those $M$-values for which the two associated $z$-roots coalesce.
Setting $\mathrm{Disc}_z[g]=0$ and solving the two resulting
quadratic factors give four branch-point values
%
%
\begin{align}
M_{\mathrm{br}}^{(s,\eta)} =
\frac{-\delta_s+\eta\sqrt{\delta_s^2+4t^2}}{2t},
\qquad
s,\eta\in\{\pm1\},
\label{eq:Mbr}
\end{align}
with $\delta_\pm=2(\Gamma\pm\ii t')$.
The indexing adopted throughout this work follows
$(s,\eta)=(+1,+1)$ for $n=1$,
$(+1,-1)$ for $n=2$,
$(-1,+1)$ for $n=3$, and
$(-1,-1)$ for $n=4$.

The second set of singular points entering the winding-number construction are the eigenvector-degeneracy points.
They arise when two distinct complex momenta ($z_1\neq z_2$) produce the same eigenvector ratio $M$ at a common energy, which implies
\begin{equation}
\det\!\left[H(z_1)-H(z_2)\right]=0.
\label{eq:det_condition}
\end{equation}
Considering the energy-free polynomial $g(M,z)=0$ and that both $z_1$ and $z_2$ will satisfy the same quadratic equation $g(M,z)=0$, together with Vieta's relations we obtain
\begin{equation}
q\equiv z_1z_2=
-\frac{1}{M^2}.
\end{equation}
Substituting this productinto Eq.~\eqref{eq:det_condition} and simplifying yields
\begin{equation}
m^2\left(1-\frac{1}{q}\right)^2
+\frac{t'^2}{q}
=0.
\end{equation}  
which can be expanded to 
\begin{equation}
m^2\left(M^2+1\right)^2
-t'^2M^2
=0.
\label{eq:quadFacs}
\end{equation}

Treating this as a quadratic equation in $M^2$ gives the four
eigenvector-degeneracy points, written using the same $(\sigma,\eta)$
parametrization as Eq.~\eqref{eq:Mdegmain} in the main text,
\begin{equation}
M_{\mathrm{deg}}^{(\sigma,\eta)}
=
\frac{\sigma t'+\eta\sqrt{t'^2-4m^2}}{2m},
\qquad
\sigma,\eta\in\{\pm1\},
\label{eq:Mdeg}
\end{equation}
with the indexing $n=1,\dots,4$ following the same convention as for
the branch points: $(\sigma,\eta)=(+1,+1)$ for $n=1$, $(+1,-1)$ for
$n=2$, $(-1,+1)$ for $n=3$, $(-1,-1)$ for $n=4$.
The indexing of $M_{\mathrm{br}}^{(n)}$ and
$M_{\mathrm{deg}}^{(n)}$ introduced above has been chosen only as an algebraic enumeration of the four solutions in each case. They do not specify whether a given point belongs to the $M_+$ or $M_-$ branch. Their correspondence with the continuously tracked curves must be established using the same analytic continuation convention used to define the curves in
Eq.~\eqref{eq:Mcurves}. Consequently, throughout this work, the branch-point and degeneracy-point pairings entering the winding-number construction are determined only through the analytically continued sheets rather than from the algebraic indexing of the roots.

\begin{figure}[t]
    \centering
    \includegraphics[width=\linewidth]{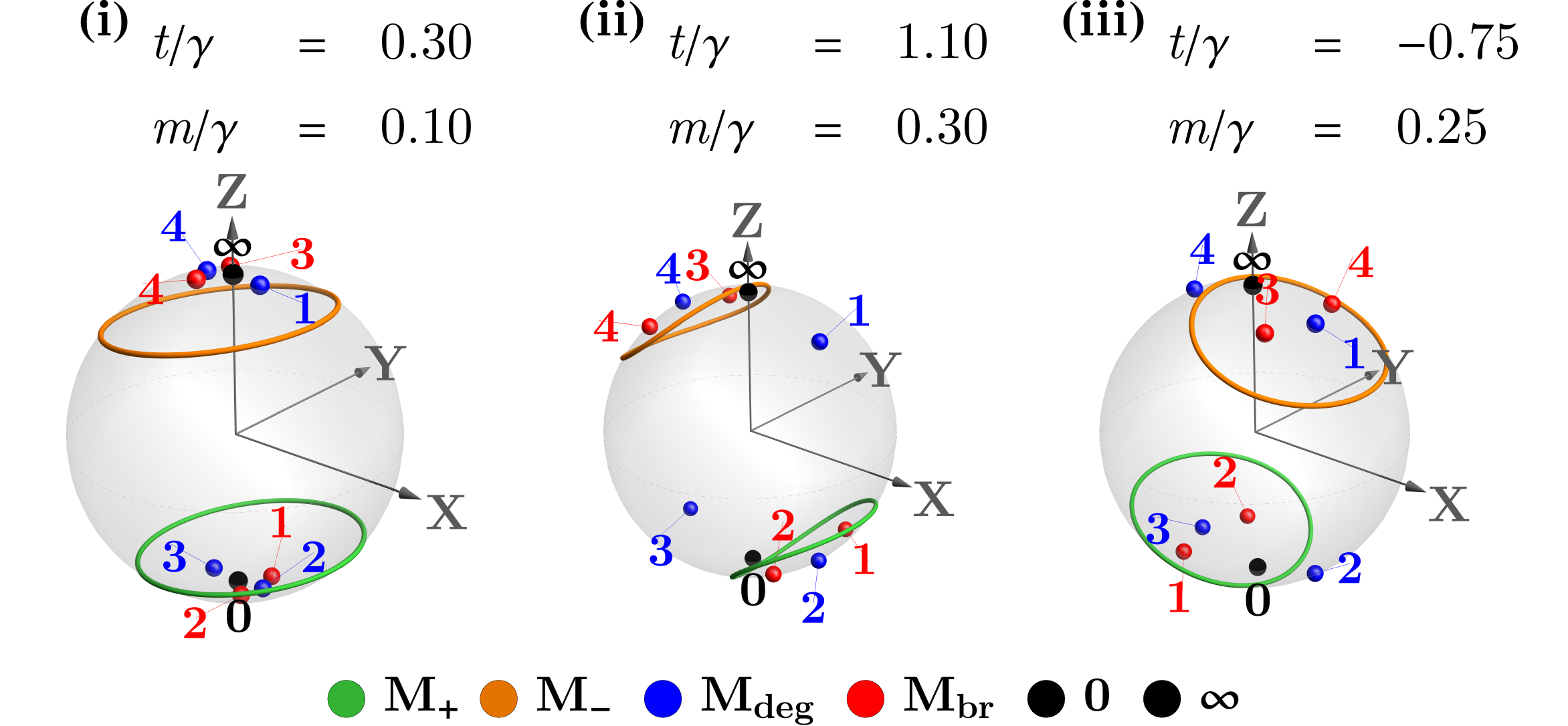}
    \caption{
    \textbf{$M$-Riemann sphere representation for the points marked in
    Fig.~\ref{fig:combined_winding}.}
    $M_+(z)$ (green) and $M_-(z)$ (orange) trajectories on the
    $M$-Riemann sphere for the three-parameter points used in
    Fig.~\ref{fig:combined_winding}(i)--(iii):
    $(t/\gamma,m/\gamma)=(0.30,0.10)$ (left),
    $(1.11,0.30)$ (middle), and $(-0.75,0.25)$ (right).
    Blue and red dots mark $M_{\mathrm{deg}}$ and $M_{\mathrm{br}}$
    respectively;the labels $n=1,\ldots,4$ follow the algebraic enumeration introduced in Appendix~\ref{app: winding number}. Black dots mark $M=0$ (south pole, $\psi=(1,0)$) and $M=\infty$ (north pole, $\psi=(0,1)$).
    The relative winding of the two curves around $M_{\mathrm{deg}}$ and $M_{\mathrm{br}}$ determines $W_j$ in Eq.~\eqref{eq:WdegWbr}.}
    \label{fig:Mriemann_appendix}
\end{figure}

Figure~\ref{fig:Mriemann_appendix} illustrates the final outcome of the $M$-Riemann construction~\cite{zhong2025topological} as applied to the present model and developed throughout this Appendix. For the three representative parameter points of Fig.~\ref{fig:combined_winding}(i)--(iii), the analytically continued trajectories $M_\pm(\ee^{\ii k})$, together with the branch points $M_{\mathrm{br}}$ and eigenvector-degeneracy points $M_{\mathrm{deg}}$, provide a complete geometric representation of the winding-number construction. Their relative arrangement completely encodes the topology of the system, from which the winding numbers in Eq.~\eqref{eq:WdegWbr} follow directly, establishing the correspondence between the analytic structure of the eigenvector mapping and the bulk topological phase.


\bibliography{bibliography}

\end{document}